# StochKit-FF: Efficient Systems Biology on Multicore Architectures

Marco Aldinucci    Andrea Bracciali    Pietro Liò
Anil Sorathiya    Massimo Torquati

# StochKit-FF: Efficient Systems Biology on Multicore Architectures

Marco Aldinucci [*]    Andrea Bracciali[†]    Pietro Liò[‡]

Anil Sorathiya[§]    Massimo Torquati[¶]

July 1, 2010

**Abstract**

The stochastic modelling of biological systems is an informative, and in some cases, very adequate technique, which may however result in being more expensive than other modelling approaches, such as differential equations. We present StochKit-FF, a parallel version of StochKit, a reference toolkit for stochastic simulations. StochKit-FF is based on the FastFlow programming toolkit for multicores and exploits the novel concept of selective memory. We experiment StochKit-FF on a model of HIV infection dynamics, with the aim of extracting information from efficiently run experiments, here in terms of average and variance and, on a longer term, of more structured data.

## 1 Introduction

The immune system is an example of a complex system formed out of its intercellular and intracellular components, which organise in space and time the immune response to pathogens through a system of regulatory nested feedbacks, both positive and negative. The modelling of part of the immune response to HIV infection is a paradigmatic scenario illustrating the challenges that computer-based modelling and analysis present for this class of problems. The immune system can be modelled by deterministic equations or by stochastic modelling approaches. Differential equations (ODEs) are effective in characterizing the system dynamics when the molecular copy number of each species is sufficiently large. Whenever the number of molecules considered is small, then a stochastic model is much more accurate on a mesoscale. The numerical solvability of stochastic models is limited to pretty small dimensions (e.g. number of species) due to their exponential complexity. The behaviour of larger systems can be described by stochastic simulations, such as those based on the *Gillespie's algorithm*, a Monte Carlo method simulating the system dynamics step by step. These methods are more accurate than the deterministic ones for describing many phenomena, however they can be highly demanding in terms of computational power, for instance when many simulations are considered for precisely characterising the overall behaviour of the system. Stochastic methods represent the most challenging methodological areas of system biology and will play a growing role in modelling, amongst the others, immune system responses to pathogens.

---

[*]Computer Science Department, University of Torino, Italy
[†]ISTI - CNR, Italy
[‡]Computer Laboratory, Cambridge University, UK
[§]Computer Laboratory, Cambridge University, UK
[¶]Computer Science Department, University of Pisa, Italy

We here illustrate the use of parallelism for supporting efficient and informative stochastic analysis of one such model. Multiple simulations exhibit a natural independence that would allow them to be treated in an *embarrassingly parallel* fashion. However, this is not possible whenever the results need to be concurrently combined or compared. In many cases, recombination is done in a post-processing phase as a sequential process whose cost in time and space depends on the number and the size of the simulation results and might be comparable to the cost of the simulation phase. A second important aspect regards execution performance: independent simulations exhibit good parallel scalability only if executed onto truly independent platforms (such as multicomputers, cluster or grid), but they might exhibit serious performance degradation if run on multicore due to the concurrent usage of underlying resources. This effect is particularly significative for I/O-bound applications since typically I/O and memory buses are shared among cores.

In this paper we introduce StochKit-FF, a parallel version of the popular StochKit framework [12], aiming at supporting the execution of multiple simulations and at combining their results on cache-coherent, shared memory multicores. This kind of architectures has been recently embraced by the whole computer industry and they equip a large range of computing platforms. StochKit-FF has been designed and developed as a low-effort, efficient porting of the StochKit main simulation loop by means of the FastFlow C/C++ programming framework, which supports efficient applications on multicore. FastFlow makes it possible to run multiple parallel stochastic simulations and to combine their results. This relies on *selective memory*, a novel data structure we designed to perform the online *alignment* and *reduction* of multiple streams of simulation results: different data streams are aligned according to simulation time and combined together according to a user-defined function, e.g. the average or others. By discussing the HIV case-study, we intend to show that this framework represents an efficient way for running multiple simulations and a promising direction for the development of effective modelling techniques, currently centered on distilling averaged values, but also more structured and informative data types on a longer term project.

Next, we summarise aspects of the immune response to HIV, recapitulate how stochastic simulations work and introduce our model. Then we overview Fast-Flow and present StochKit. A section on experimental results follows before some concluding remarks.

## 2  A stochastic model of the immune response to HIV

Ordinary differential equations (ODEs) based models have long been used for immune system and viral infection modeling [11, 15, 14]; they focus on the average behavior of large populations of identical objects; large complex system of nonlinear equations need often be solved numerically. When considering a small number of molecules, which is highly probable if we consider immune cell interactions in a small volume or if the reactions occur far from thermodynamic equilibrium, or when considering randomness and irregularities found at all levels of life, such as the random motion of cellular components, then a stochastic model is much more accurate on a mesoscale. For instance, the master equation is a differential-difference equation for the time-dependent probability density of the number of molecules of each species. Stochastic methods are those based on the *Gillespie's algorithm*, a Monte Carlo method simulating the reactions step by step [8]. Such stochastic methods are more effective than the deterministic ones to describe the above mentioned irregularities and also crucial chemical reactions.

We adopt an agent-based approach: simulation considers the actions of a large number of simple entities, or agents (viruses and cells), in order to observe emerging properties from their interaction, which is based on local rules [6].

Briefly, each agent consists of state variables and a set of rules that govern its behavior. Agents can interact either directly with each other or indirectly through the environment. Agent behaviour consists of actions, e.g. cellular interactions, that cause a state transition of the modelled system, e.g. a variation in the amount of agents. Actions are stochastic in the sense that their occurrence in time has an associated probability. Associated probability distributions are generally memoryless, typically negative exponential distributions with the rate as parameter, and hence the overall system behaviour can be interpreted as a Continuous Time Markov Chain (CTMC). Systemic emergent properties can be sensitive to the local presence of minimal (integer) quantities of agents/molecules/cells [18]. The combined behavior of these agents is observed in a discrete-time or event-driven stochastic simulation, from given initial conditions. Given the current state of the system, a single transition amongst the possible ones is selected, and the state updated accordingly. The Gillespie's algorithm [8] determines the next transition and the time at which it occurs, exactly according to the given probability distributions. Each so-determined possible evolution of the system is called a *trajectory*. Long time and large computing resources may be required in order to correctly evaluate parameters and determine fluctuations and averages of the system behaviour

| | | | | |
|---|---|---|---|---|
| $U_0 \xrightarrow{\lambda} U_0 + U$ | $T + F \xrightarrow{\delta_{tf}} F$ | $I_5 \xrightarrow{\sigma_2} I_5 + Z_5$ | $V_5 \xrightarrow{\mu} V_4$ | $Z_5 \xrightarrow{\delta_z} 0$ |
| $U + F \xrightarrow{\delta_{uf}} F$ | $T + V_5 \xrightarrow{\beta_{R5}} I_5$ | $I_4 \xrightarrow{\sigma_2} I_4 + Z_4$ | $I_5 \xrightarrow{\pi} V_5 \times 200$ | $Z_4 \xrightarrow{\delta_z} 0$ |
| $U \xrightarrow{\delta_{ut}} T$ | $T + V_4 \xrightarrow{\beta_{X4}} I_4$ | $I_5 + Z_5 \xrightarrow{\sigma_1} U + I_5 + Z_5$ | $I_4 \xrightarrow{\pi} V_4 \times 200$ | $V_5 \xrightarrow{\delta_v} 0$ |
| $U \xrightarrow{\delta_{uz}} Z_4$ | $Z_5 + V_5 \xrightarrow{\gamma_{R5}} I_5$ | $I_4 + Z_4 \xrightarrow{\sigma_1} U + I_4 + Z_4$ | $V_4 \xrightarrow{\sigma_3} F + V_4$ | $V_4 \xrightarrow{\delta_v} 0$ |
| $U \xrightarrow{\delta_{uz}} Z_5$ | $Z_4 + V_4 \xrightarrow{\gamma_{X4}} I_4$ | $I_5 + Z_5 \xrightarrow{\delta_{iz}} Z_5$ | $I_5 \xrightarrow{\delta_i} 0$ | |
| $T \xrightarrow{\delta_t} 0$ | | $I_4 + Z_4 \xrightarrow{\delta_{iz}} Z_4$ | $I_4 \xrightarrow{\delta_i} 0$ | |

Table 1: Reaction-centric view of the HIV model.

when simulating trajectories.

## 2.1 HIV and the immune response in formulae

The model we are drafting in this section follows those in [11, 16] and is sketched as simply but meaningfully as possible, given its purposes here. During the HIV infection multiple strains of the virus arise, we consider two phenotype classes, $V5$ and $V4$. They allow virus entry in cells through two membrane receptors, CCR5 and CXCR4, respectively. The mutation from $V5$, initially prevailing, to the more aggressive $V4$ has been correlated with a faster progression of the disease to the AIDS phase. The immune response is based on the action of several cells, some of which strain specific ($T, Z5$ and $Z4$), which can also be infected by the viruses. We also consider the Tumor Necrosis Factor $F$, which induces bystander death of several cells. Infection is characterised by the progressive loss of $T$ cells, infected together with $Z5$ and $Z4$ by the virus in order to proliferate.

The set of possible reactions is reported in Table 1, according to a reaction-centric view, while the behaviour of a specific agent is determined by the union of the reactions to which it participates. Reactions are labeled with their stochastic rates and are commented below.

Mature (uninfected) $T$ cells have precursors, the immature Naive $T$ cells ($U$), produced at a constant rate $\lambda$ ($200C/\mu l t^{-1}$, with $C/\mu l$ cells per micro-liter and $t^{-1}$ the inverse of time). $U$ cells are transformed into mature $T$ cells with rate $\delta_{ut}$ ($0.129 t^{-1}$), and also into $Z4$ and $Z5$ cells with rate $\delta_{uz}$ ($0.005 t^{-1}$), and are cleared out by the interaction with $F$ (Tumor Necrosis Factor, a cytokine) with rate $\delta_{uf}$ ($10^{-5} \mu l/C t^{-1}$). $T$ cells are cleared with rate $\delta_t$ ($0.01 t^{-1}$).

The increase of $F$ causes the death of the $T$ cells at rate $\delta_{tf}$ ($10^{-5} \mu l/C t^{-1}$). Viral strains $V5$ and $V4$ interacting with $T$ cells produce infected cells $I5$ and $I4$ at rate $\beta_{R5}$ and $\beta_{X4}$ respectively. Similarly, $V5$ and $V4$ interacting with $Z5$ and $Z4$ cells produce infected cells at rate $\gamma_{R5}$ and $\gamma_{X4}$, respectively. The strains $Z5$ and $Z4$ are cleared at rate $\delta_z$ ($0.01 t^{-1}$).

$Z$ cells proliferate due to infection in the system with the proliferation rate $\sigma_2$ ($0.001 \mu l/C t^{-1}$). Moreover, the production of $U$ cells increases due to the presence of $Z$ and infected cells at rate $\sigma_1$ ($0.0001 \mu l/C t^{-1}$). This is a simplifying working hypothesis: the effects of the presence of $Z$ and $I$, i.e. the increase of $U$, are modelled rather than the actual mechanisms that lead to the generation of new $U$s (via the dendritic cells signalling to naive T cells for optimal activation [9]). The infected cells are killed by $Z$ cells with rate $\delta_{iz}$ ($0.06 \ \mu l/C \ t^{-1}$).

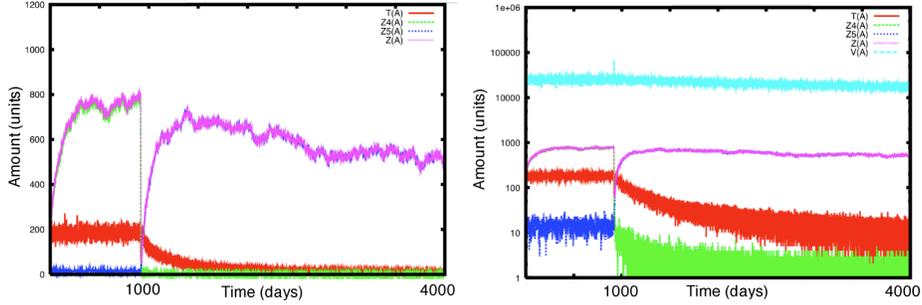

Figure 1: HIV Model: a single trajectory. *1. (left)* The immune cell dynamics is shown over a time span of 4000 days. It is evident both the "noisy" nature of cell time courses and the $V5 - V4$ mutation at about day 1000. Note how the most aggressive strain $V4$ defeats $T$ cells (An amount of T cells less than about 200 is considered the threshold for AIDS). *2. (right)* Also the virus $V5 + V4$ is reported on a log-scale, which highlights the effects of the mutation on the virus (note the high peak perturbation) and coincidently on the immune system, and also the dynamics for small amounts of cells.

Mutation of $V5$ strains into $V4$ strains happens at rate $\mu$ ($2.5 \times 10^{-3}$). The bursting of infected cells produces about 200 viruses at rate $\pi$ ($200\mu l/Ct^{-1}$). The accumulation of $F$ is proportional to the total amount of $V4$ viruses, and happens at a rate $\sigma_3$ ($0.0001t^{-1}$). The infected cells are cleared at rate $\delta_i$ ($0.33t^{-1}$). Viruses are cleared at rate $\delta_v$ ($2t^{-1}$).

The parameters used have been referred from literature, e.g. [11, 15, 14, 17, 13] and sometimes tuned against known aspects of the macroscopic behaviour of the system. The initial conditions for the simulations of our system are: $U0 = 1, U = 200, T = 1000, Z4 = 250, Z5 = 250, I4 = 0, I5 = 0, V4 = 0, V5 = 100, F = 0$. The infectivity of $V5$ and $V4$ depends on the concentration of TNF in the system. Initially, in the absence of $F$, the infectivity rate is constant, $\beta = 4 \times 10^{-5} \mu l/C \ t^{-1}$ for both $V5$ and $V4$. When $V4$ appears, $\beta_{R5} = \beta - K * F$ decreases and $\beta_{X4} = \beta + K * F$ increases because of the amount of $F$ ($K = 10^{-7}(\mu l/C)^2 t^{-1}$ and $0 \leq \beta_{R5} \leq \beta$). Analogously: $\gamma_{R5/X4} = \gamma \mp K * F$ ($\gamma = 2 \times 10^{-5} \mu l/C \ t^{-1}$).

It is worth pointing out, as a modelling methodology, that we embedded a notion of time in the model by means of a stochastic event expected to occur *at about* a given time instant and used to trigger the $V5$ to $V4$ mutation at a certain stage of the disease progression.

Figure 1 reports a trajectory of our model. Next section will introduce the framework that allowed it to be computed.

# 3 Parallel Stochastic Simulations

In Monte Carlo simulations, each individual trajectory represents just one possible way in which the system might have reacted over the entire simulation time-span. Many trajectories might be needed to get a representative picture of how the system behaves on the whole. Processing and combining many trajectories may lead to very high compulsory cache miss-rate and thus become a memory-bound (and I/O-bound) problem. This in turn may require a huge amount of storage space (linear in the number of simulations and the observation size of the average trajectory) and an expensive post-processing phase, since data should be retrieved from permanent storage and processed. Eventually, the computational problem hardly benefits from the latest commodity multi-core architectures. These architectures are able to exhibit an almost perfect speedup with independent CPU-bound computations, but hardly replicate such a performance for memory-bound and I/O-bound computations, since the memory is still the real bottleneck of this kind of architectures.

Tackling these issues at the low-level is often unfeasible because of the complexity of the code and of the need to keep the application code distinct from platform-specific performance tricks. Typically, low-level approaches only provide the programmers with primitives for flow-of-control management, synchronisation and data sharing. In this regard, the shared memory paradigm, which is natively supported by multicores, is often and incorrectly considered simpler with respect to message passing paradigm. While writing a first working prototype is certainly faster, tuning it for performance is often a nightmare due to the non-trivial effects induced by memory fences (used to implement mutex) on data replicated in core's caches.

Designing suitable high-level abstractions for parallel programming is a long standing problem [5]. Recently, high-level parallel programming methodologies are receiving a renewed interest, especially in the form of pattern-based programming [7, 10, 4]. FastFlow belongs to this class of programming environments.

## 3.1 The FastFlow Parallel Programming Environment

FastFlow is a parallel programming framework aiming at *simplifying* the development of *efficient* applications for multicore platforms, being these applications either brand new or ports of existing legacy codes. The key vision underneath FastFlow is that effortless development and efficiency can both be achieved by raising the level of abstraction in application design, thus providing designers with a suitable set of parallel programming patterns that can be compiled onto efficient networks of parallel activities on the target platforms. To fill the abstraction gap, as shown in Figure 2, FastFlow is conceptually designed as a stack of layers that progressively abstract the shared memory parallelism at the level of cores up to the definition of useful programming constructs and patterns.

At the lowest tier of the FastFlow system we have the architectures that it targets: cache-coherent multiprocessors, and in particular commodity homoge-nous multicore (e.g. Intel core, AMD K10, etc.).

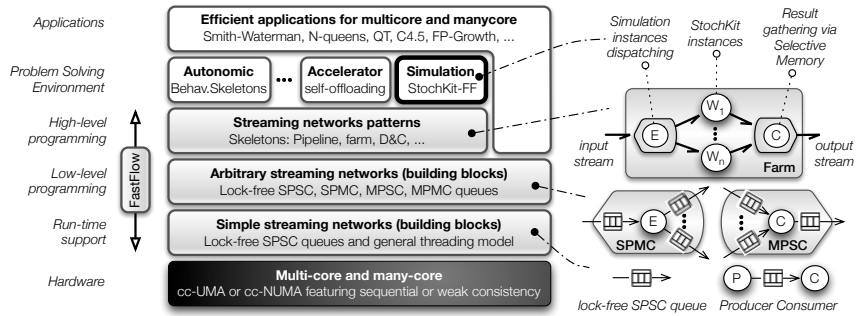

Figure 2: FastFlow layered architecture with pattern examples.

The second tier provides mechanisms to define simple streaming networks whose *run-time support* is implemented through correct and efficient lock-free Single-Producer-Single-Consumer (SPSC) queues. This kind of queues do not require (under mild requirements [1]) any lock or memory barrier, and thus they constitute a solid ground for a low-latency synchronization mechanism for multicore. These synchronizations, which are asynchronous and non-blocking, do not induce any additional cache invalidation as it happens in mutual exclusion primitives, and thus do not add a significant overhead. The third tier generalises one-to-one to one-to-many (SPMC), many-to-one (MPSC) and many-to-many (MPMC) synchronizations and data flows, which are implemented using only SPSC queues and arbiter threads. This abstraction is designed in such a way that arbitrary networks of activities can be expressed while maintaining the high efficiency of synchronisations.

The next layer up, i.e., *high-level programming*, provides a programming framework based on parallelism exploitation patterns (a.k.a. *skeletons* [5]). They are usually categorised in three main classes: Task, Data, and Stream Parallelism. FastFlow specifically focuses on Stream Parallelism, and in particular provides: *farm*, *farm-with-feedback* (i.e. Divide&Conquer), *pipeline*, and their arbitrary nesting and composition. These high-level skeletons are actually factories for parametric patterns of concurrent activities, which can be instantiated with sequential code or other skeletons, then cross-optimised and compiled together with lower FastFlow tiers. The skeleton disciplines concurrency exploitation within the generated parallel code: the programmer is not required to explicitly interweave the business code with concurrency related primitives.

The set of skeletons provided by FastFlow could be further extended by building new C++ templates or even further abstracted to derive problem specific skeletons. We refer to [3] for any further details. FastFlow is open source available at http://sourceforge.net/projects/mc-fastflow/ under LGPL3.

## 3.2 Parallel StochKit: StochKit-FF

StochKit [12] is an efficient, extensible stochastic simulation framework developed in the C++ language. It aims at making stochastic simulation accessible to practicing biologists and chemists, while remaining open to extension via new stochastic and multi-scale algorithms. It implements the popular Gillespie algorithm, explicit and implicit tau-leaping, and trapezoidal tau-leaping methods.

StochKit-FF extends StochKit v1 with two main features: The support for the parallel run of multiple simulations on multicores, and the support for the online (parallel) *reduction* of simulation results, which can be performed according to one or more user-defined associative and commutative functions. StochKit v1 is coded as a sequential C++ application exhibiting several non-reentrant functions, including the random number generation.[1] Consequently, StochKit-FF represents a significative test bed for the FastFlow ability to support parallelisation of existing complex codes. The parallelisation is supported by means of high-level parallel patterns, which could also be exploited as *parametric code factories* to parallelise existing, possibly complex C/C++ codes [1].

In particular, StochKit-FF exploits the FastFlow *farm* pattern, which implement the functional replication paradigm: a stream of independent data items are dispatched by an *Emitter* thread to a set of independent *Worker* threads. Each worker produces a stream of results that is gathered by a *Collector* thread into a single output stream. The *farm* paradigm is sketched in Figure 2.

In StochKit, a simulation is invoked by way of the `StochRxn()`, which realises the main simulation loop. The propensity function and initial conditions are among its parameters. StochKit-FF provides the programmer with `StochRxn_ff()` function, which has a similar list of parameters, but invokes a parametric simulation modelling either a number of copies of the same simulation or a set of parameter-sweeped simulations. `StochRxn_ff()` embodies a *farm*: the emitter unrolls the parametric simulation into a stream of standard simulations (represented as C++ objects) that are dispatched to workers. Each worker receives a set of simulations, which are sequentially run by way of the `StochRxn()`, which is basically unchanged[2] with respect to the original StochKit. Each simulation produces a stream of results, which are locally *reduced* within each worker into a single stream [5]. The collector gathers all worker streams and *reduces* them again into a single output stream. In the case it is required, the worker can directly write the full simulation results of all simulations into different files.

Overall, the parallel *reduction* happens in a systolic (tree) fashion via the so-called *selective memory* data structure.

---

[1] At the time of writing, a new beta multithreaded version of StochKit has been made available. However, also this new version does not offer the possibility to combine trajectories from different simulations, which are written into different files.

[2] It is made thread-safe by substituting the *sprng* random generator with *boost*.

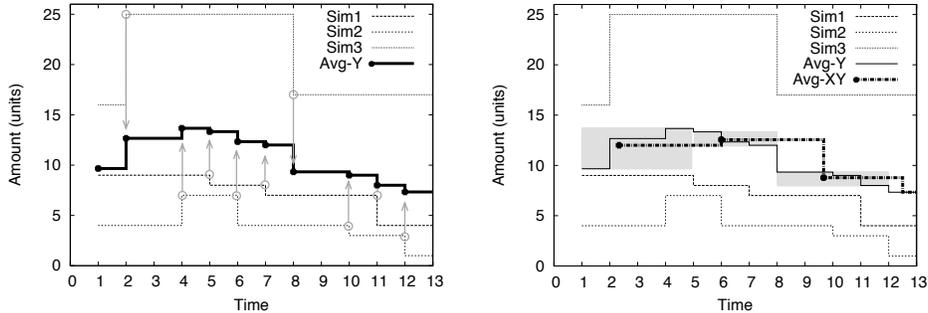

Figure 3: Selective Memory with average. *1. (left)* Curve Avg-Y is derived via oversampling and time-aligned reduction (average along $Y$ axis) of $k$ independent simulations (arrows highlight oversampling). *2. (right)* Avg-XY is derived by the reduction (average along $X$ axis) of $k$ successive points of Avg-Y (grey boxes highlight averaging zone).

### 3.2.1 Selective Memory

Together with StochKit-FF, we introduce the *selective memory* concept, i.e. a data structure supporting the on-line *reduction* of time-aligned trajectory data by way of one or more user-defined associative and commutative functions. Selective memory distinguishes from standard parallel *reduce* operation [5, 2] because it works on (possibly unbound) streams, and aligns simulation points (i.e. stream items) according to simulation time before reducing them: since each simulation proceed at a variable time step, simulation points coming from different simulations cannot simply be reduced as soon as they are produced. Selective memory behaves as an array of buffers that abstractly realise a sliding window that follows the wavefront of generated simulation points along simulation time. It keeps the bare minium amount of simulation points from different simulation to produce a slice of aligned simulation points, which are popped from selective memory as soon as they are reduced.

The behaviour of selective memory is exemplified in Figure 3 using average as combining function. Simulation points from different simulations are first averaged at aligned simulation time points: such computed average results oversampled with respect to single simulations (Figure 3 left). This oversampling is possibly reduced by applying the same technique along time axis (Figure 3 right). Overall, selective memory produces a combined simulation that has been adaptively sampled: time intervals exhibiting a higher variability across different simulations exhibit an higher sampling rate. Selective memory effectively mitigates the memory pressure of result logging when many simulations are run on a multicore, because it substantially reduces the output size, and thus capacity misses and the memory bus pressure.

**The StochKit-FF implementation of the HIV model.** StochKit-FF main-tains the same interface of StochKit. A model is specified by providing C++ files containing the needed information in suitable data structure. In our case: *i)* a list of constant definitions for all the *parameters* used; *ii)* an integer *state vector* with an entry for each species in the model, initialised with the initial conditions for the simulation; *iii)* an integer *stoichiometric matrix* with an entry $i, j$ for the amount of the $i$-th species involved in the $j$-th reaction, e.g. $-1$ for a consumed species or 200 for the viruses released by $I_4 \xrightarrow{\pi} V_4 \times 200$ (Table 1); and *iv)* the *propensity vector* associating to each reaction its propensity, then employed by the Gillespie algorithm in the selection of the next reaction.

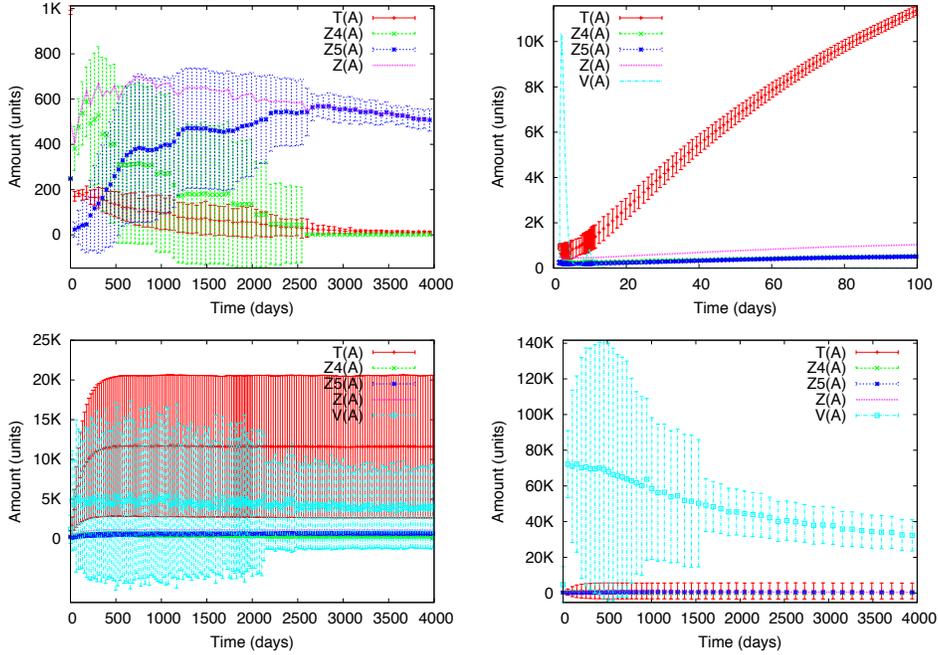

Figure 4: HIV Model: average and variance for multiple trajectories (16x). *Left to right:* *1.* A focus on the immune response. *2-4* Sensitivity analysis for $\delta_t = 5.0, 3.0$ and $0.5$.

In our case, reactions follow the law of mass action and depend on the amount of reactants and the associated rates, intuitively the more the reactants (and the rate) the highest the probability of being selected, e.g. $I_4 \times \pi$ for the previous reaction. Finally, the desired duration of the simulation interval has to be specified.

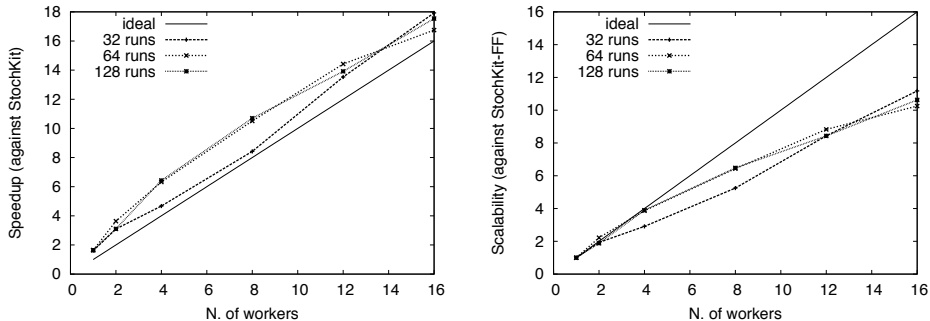

Figure 5: *1. (left)* Speedup of StochKit-FF against StochKit. *2. (right)* Scalability of StochKit-FF(n) against StochKit-FF(1), where n is the number of worker threads.

## 4 Experiments and Discussion

We report on some experiments performed on the `ness.epcc.ed.ac.uk` platform (Sun X4600 SMP - 8 x Dual-Core AMD Opteron 1218, 32 Gb memory) hosted at EPCC, University of Edinburgh.

Figure 4 (top-left) is a focus on the immune response averaged over 16 simulations. The averaged amounts of $Z4$, $Z5$, their sum $Z$, and $T$ are reported together with a representation of their variance (obtained by rough summation). The variance of $Z4$ and $Z5$ is large till 2500 days, showing tight coupling i.e. interdependence. Then, the variance of T decreases continuously, while the one of $Z5$ decreases together with the (averaged) amount of $Z5$, as it is not much involved in dynamics after the mutation to $V4$, while the number of the budding viruses still have high probability of infecting and killing cells.

The other three pictures in Figure 4 describe a sensitivity analysis for $\delta_t$, the parameter regulating the lifespan of the virus and originally set to 1.5. By a large analysis of the model, $\delta_t$ has resulted in being very influential on the emerging behaviour of the system as it strongly impacts on the diffusion of the infection. In *2* the virus is immediately cleared out (the picture is relative to the interval [1,100]), $T$ rapidly increases, and the system is very stable, with a low variance. In *3* the immune response still prevails, but the system appears much perturbed. In *4*, well below the standard value of $\delta_t$, the virus clearly prevails. Initially variance is high, over large numbers for the viral load, then it tends to stabilise towards a steady state.

The performances of StochKit-FF have been evaluated on multiple runs of the HIV case-study. A single run of the simulation with StochKit averagely produces $\sim$ 150M simulation points for 4000 days of simulated time, which will turn into about 6 GBytes of generated data; multiple runs of the same simulation will need a linearly greater time and space. These simulations can be naively parallelised on a multicore platform by running several independent instances, which however, will compete for memory and disk accesses, thus lead to suboptimal performances in the case of high output pressure. Observe, as an example, that the HIV simulation with a 1:10 sampling (outputting 1 simulation point every 10) spends 64 seconds for the simulation and 133 seconds in the output phase on our testing platform. An additional linear time (at least) in the number and the size of outputs should be spent in the postprocessing phase for the recombination of results (6 MBytes per simulation per the number of runs).

StochKit-FF mainly attacks these latter costs by online reducing the outputs of simulations, which are run in parallel. In Figure 5, average and variance are used as combining functions. As shown in Figure 5 (left), StochKit-FF exhibits a superlinear speedup with respect to StochKit in all tested cases (32, 64, 128 repetitions of the HIV simulation). This superlinear speedup is mainly due to the fact that StochKit-FF is about two times faster than StochKit even when running with just one thread. They are mainly due to FastFlow memory allocator that is faster than standard memory allocator on the tested platform, and some little code optimisation involving memory management that has been introduced during the port. As shown in Figure 5, StochKit-FF exhibits a good scalability also when compared with the sequential (one-thread) version of StochKit-FF.

## 5 Concluding remarks

We have introduced StochKit-FF, developed within FastFlow, which has supported the porting of a complex application with minimal modifications required. StochKit-FF allows multiple stochastic simulations of biological systems to be run efficiently and their results to be suitably combined according to the user needs. For the aims of recombinations, we have developed the concept of selective memory. For the development and for this presentation we have grounded StochKit-FF on a realistic model of HIV infection, as the stochastic analysis of this kind of systems is playing a growing role in systems biology. We have presented simulations and experiments for this model, highlighting aspects of the emerging behaviour of the system under different assumptions. As discussed, performances are convincing in terms of efficiency, close to optimality, and scalability. We are interested in further improving performances, in developing more informative recombining functions, which will allow more structured data to be extracted from multiple simulations, and in testing the framework on different domains.

## Acknowledgments

Marco Aldinucci and Andrea Bracciali have been partially supported by a grant of the HPC-Europa 2 *Transnational Access* programme. Marco Aldinucci has been partially supported by the *BioBITs* ("Developing White and Green Biotechnologies by Converging Platforms from Biology and Information Tech-nology towards Metagenomic").